\documentclass{article}
\usepackage{icrctc07}
\def\diffunits{\mbox{GeV cm}^{-2}\,\mbox{s}^{-1}\,\mbox{sr}^{-1}}
\title{Particle acceleration in relativistic subluminal shock environments}
\shorttitle{Particle acceleration in relativistic subluminal shock environments}
\authors{A. ~Meli$^a$, J.~K.~Becker$^b$, J.~J.~Quenby$^c$ and J.~L\"unemann$^b$}
\shortauthors{A.~Meli and et al.}
\afiliations{
(a) Department of Physics, National and Kapodistrian University of Athens, 15783 Zografos, Athens Greece\\
        (b) Institut f\"ur Physik, Dortmund Universit\"at,  44221 Dortmund Germany \\
        (c) Astrophysics Group,  Blackett Laboratory, Imperial College London,
        SW7, 2BW, London UK
}
\email{ameli@phys.uoa.gr}
\abstract{
The understanding of the particle spectra resulting from acceleration in
relativistic shocks as they occur in extragalactic sources, is essential
for the interpretation of the cosmic ray spectrum above the ankle
($E_p>3\cdot 10^{18}$~eV). It is believed that extragalactic sources like
Active Galactic Nuclei and Gamma Ray Bursts can produce particle spectra
up to $E_p\sim 10^{21}$~eV.
In this contribution, subluminal shocks are investigated with
respect to different shock boost factors $\Gamma$ and the inclination
angle between the shock normal and the magnetic field $\psi$. A
correlation between the boost factor and the spectral behavior of the
emitted particles is found. The results are compared to Active Galactic Nuclei and
Gamma Ray Burst diffuse cosmic ray contribution and the observed cosmic ray spectrum at 
the highest energies.
}

\begin{document}
\maketitle

\section{Introduction}

The observations of electron synchrotron radiation in the radio regime 
and the X-ray variations from astrophysical environments such as Super Novae (SN), 
Active Galactic Nuclei (AGN) and Gamma Ray Bursts (GRBs), indicate that there is a mechanism with which 
primary particles (protons or electrons) gain large amounts of energy via
acceleration. Connected to this implied mechanism, it is accepted that shocks are the source of the 
acceleration of the observed cosmic rays.
The work of the late 70s by a number of authors such as \cite{Krymsky77,Bell78a,Bell78b,Axford78,BlandOst78} 
established the basic mechanism of particle diffusive acceleration in non-relativistic shocks. 
Later on studies on relativistic shocks appeared (e.g. \cite{Peacock,KirkSchneider,QuenbyLieu,Ellisonetal})
and questioned the nature and the limits of the relativistic diffusive acceleration versus the non-relativistic 
shock mechanism. In this work, we present results on relativistic particle acceleration which
are extended to relativistic source models (i.e.~AGN, GRBs). These involve
subluminal shocks with plasma shock velocity ranging from Lorentz boost factors, 
$\Gamma=1 \rightarrow 1000$, using pitch angle and magnetic
scattering. In the first section, the Monte Carlo simulations are described,
while in the second section the results are presented for different boost
factors. In the third section, the Monte Carlo spectra are used to calculate
the possible contribution of AGN and GRBs to the observed spectrum of charged
cosmic rays. The final section summarizes the results.

\section{Numerical simulations}

The Monte Carlo codes simulate relativistic oblique shocks with application to
relativistic environments of AGN and GRBs.
Initially the particles are considered relativistic, isotropically injected upstream of the shock and 
they are let to move towards the shock and along the way they collide with the
scattering centers. Consequently, as they keep scattering between the upstream and
downstream regions of the shock (its width is much smaller than the particle's gyroradius),
they gain a considerable amount of energy.
During our simulations, Lorentz relativistic transformations are performed from the local plasma frames to the 
normal shock frame to check for shock crossings. \\
Particles leave the system if they escape far downstream 
at a spatial boundary $r_b$, or if they reach a defined maximum energy $E_{max}$, 
even though other physics describing particle escape or energy loss 
would probably need to be taken into account in realistic situations (forthcoming paper). The 
downstream spatial boundary required can be estimated from the solution of the convection-diffusion
equation in a non-relativistic, large-angle scattering approximation in the
downstream plasma. \\
The mean free path is calculated in the respective fluid frames by the formula $\lambda=\lambda_{\circ} \cdot p_{1,2}$, assuming a momentum dependence to this mean free path for scattering 
along the field,  related to the 
spatial diffusion coefficient, $\kappa$. In the shock normal, or in the $x$ direction, 
the diffusion coefficient is given by $\kappa_{\|}=\lambda v/3$ where $\kappa=\kappa_{\|}cos^{2} \psi$.
At the scattering centers the energy
(momentum) of the particle is kept constant and only the direction of the velocity vector ${\bf
\upsilon}$ is randomized, using a computational random number generator. 
Runs were performed with different spatial boundaries to investigate
the effect of the size of the acceleration region on the spectrum, as well as to find a region where
the spectrum is size independent. In the small-angle scattering case, the inherent anisotropy due to the
high downstream sweeping effect, may greatly modify this analytical estimate.
A splitting technique is used similar to the one used in the Monte Carlo simulations of \cite{MeliQa},
\cite{MeliQb}, so that when an energy level is reached such that only few accelerated cosmic
rays remain, each particle is replaced by a number of $N$ particles of statistical
weight $1/N$, so as to keep roughly the same number of cosmic rays being followed.
Workers on the field \cite{Galletal99} have demonstrated that small angle
scattering (pitch angle diffusion), with $\delta \theta < 1/\Gamma$, applies with $\theta$ measured 
in the upstream fluid frame for scattering in a uniform  field or a randomly oriented set of 
uniform field cells. This arises because particles attempting to penetrate upstream from the shock 
are swept back into the shock before they can move far in pitch.
For this case, we allow particles' pitch angle diffusion 
to angles chosen at random, up to an angle $\sim 10/\Gamma$, where $\Gamma$ is the 
upstream gamma, measured in the normal shock frame. \\
For the oblique shock cases that we apply here, provided the field directions encountered are reasonably isotropic in 
the shock frame, we know that $\tan\psi_{1}=\Gamma_{1}^{-1}\tan\psi_{s}\sim \Gamma_{1}^{-1}\sim \psi_{1}$
where '1' and 's' refer to the upstream and normal shock frames respectively.
Given the current interest in following shock acceleration in the relativistic flow in test particle regime,
we outline a typical approach to such computations. Using the guiding center approximation,
a test particle moving a distance, $d$, along a field line at $\psi$ to the shock normal,
in the plasma frame has a probability of collision within $d$ given by $P(d)=1-\exp(-d/\lambda)=R$.
Where the random number $R$ is $0 \leq R \leq 1$. Weighting the probability by the current in the 
field direction, $\mu$, then $d=-\lambda \mu \ln R$. Between the 'i th' and 'i+1 th' scattering, 
a shock crossing condition is checked, by finding the position and time in the normal shock frame, as
\begin{displaymath}
\Delta x_s^{i+1} =  \Gamma_1(\Delta x_1^{i}+V_1|\frac{\Delta x_1^{i}sec\psi_1}{v_1\mu_1}|)
\end{displaymath}
\begin{displaymath}
\Delta t_s^{i+1} =  \Gamma_1(|\frac{\Delta x_1^{i} sec\psi_1}{v_1\mu_1}|+V_1\Delta x_1)
\end{displaymath}
where '1' refers to the quantities that are calculated in the upstream fluid frame. For further details on 
relativistic transformation calculations see \cite{paper_in_prep}.
We note that for the description of the
physical quantities necessary throughout the simulations we use the fluid rest frames,
the normal shock frame and the de Hoffman-Teller frame where ${\bf E}$=0. 
Furthermore, as the theory of the oblique shocks postulates, \cite{Hudson65}, from the 
conservation of the first adiabatic invariant, one can find the new pitch angle in the downstream frame 
and similar transformations allow the particle scattering to be followed in this frame. 

\section{Results}

\subsection{Energy spectra}

\begin{figure}[h!]
\begin{center}
\includegraphics[width=7.5cm]{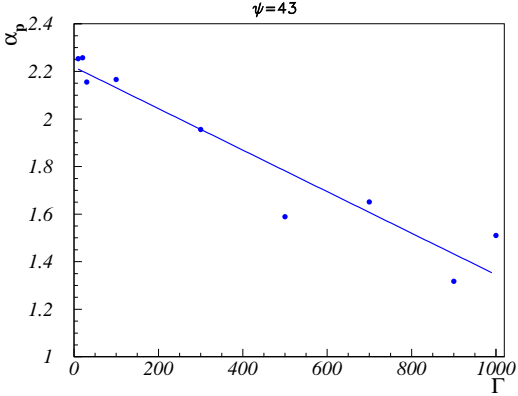}
\caption{Correlation between the shock boost factor $\Gamma$ and the spectral index, for shock inclination of $\psi=43^{\circ}$ . The correlation appears to be linear.}
\label{index_gamma_43}
\end{center}
\end{figure}
As discussed above, Monte Carlo calculations have been performed for different
parameter settings. Three different shock inclinations have been simulated,
i.~e.~$\psi=23^{\circ},\,33^{\circ}$~and~$43^{\circ}$, where $\psi$ is the
angle between the shock normal and the magnetic field. For each of the angles, nine
different boost factors have been used,
$\Gamma=10,\,20,\,30,\,100,\,300,\,500,\,700,\,900$ and~$1000$. Results
will be presented in detail in~\cite{paper_in_prep}. Furthermore, each single spectrum is
fit by a single power-law approximation, $dN/dE\propto E^{-\alpha_p}$, in order to get an estimate of
the spectral index $\alpha_p$. At the lowest energies, particle injection
features are excluded. Since AGN and GRBs are likely to accelerate particles
up to $\sim 10^{21}$~eV, this value is used as the maximum energy. This way,
each shock boost factor $\Gamma$ is associated with a spectral index
$\alpha_p(\Gamma)$. A linear correlation between $\alpha_p$ and $\Gamma$ is
found as it is displayed in Fig.~\ref{index_gamma_43}. The figure shows the
results for $\psi=43^{\circ}$, the results for $\psi=23^{\circ}$ and
$\psi=33^{\circ}$ are comparable. A linear fit is applied with a numerical
result which is comparable for all three angles,
\begin{eqnarray}
\alpha_p(\psi=23^{\circ})&=&2.1-0.7\cdot 10^{-3}\cdot \Gamma\label{23_cor:equ} \nonumber\\
\alpha_p(\psi=33^{\circ})&=&2.0-0.9\cdot 10^{-3}\cdot \Gamma\label{33_cor:equ} \nonumber\\
\alpha_p(\psi=43^{\circ})&=&2.2-0.9\cdot 10^{-3}\cdot \Gamma\label{43_cor:equ}\,.\nonumber
\end{eqnarray}
This implies three main things for relativistic, subluminal shocks:
\begin{enumerate}
\item The result is basically independent of the inclination angle between
  shock normal and magnetic field, as long as the shocks are oblique.
\item Mildly-relativistic shocks have spectral indices of around $\alpha_p\sim
  2.0-2.4$.
\item The more relativistic the shock, the flatter the
  spectrum. Highly-relativistic shocks with $\Gamma\sim 1000$ can produce spectra as flat as
  $\alpha_p\sim 1.5$.
\end{enumerate}

The above findings seem to support similar simulation results and observational evidences regarding 
irregular and flat spectra from many GRBs (further details in \cite{paper_in_prep}).

\subsection{Contribution to the charged Cosmic Ray spectrum}

In this section, the spectra of different boost factors are associated to
different source classes in order to calculate the possible contribution to
the diffuse spectrum of charged cosmic rays. Specifically, it is assumed
that the cosmic ray spectrum above the ankle, $E>10^{18.5}$~eV, is produced by
protons accelerated in AGN and GRBs sources. For AGN, a fix
boost factor of $\Gamma=10$ is used. In the case of GRBs, it is assumed that
$10\%$ of all GRBs have $\Gamma=100$, $80\%$ have $\Gamma=300$ and the
remaining $10\%$ have $\Gamma=1000$. It is further assumed that the observed
spectrum of charged cosmic rays is produced by a linear combination of AGN,
$dN_{AGN}/dE$, and GRBs, $dN_{GRB}/dE$,
\begin{displaymath}
\frac{dN_{CR}}{dE}=\alpha\cdot \frac{dN_{AGN}}{dE}+(1-\alpha)\cdot \frac{dN_{GRB}}{dE}\,, 
\end{displaymath}
with $0\leq \alpha \leq 1$. Source evolution has been taken into account by
using the source density given by the observation of X-ray
AGN~\cite{hasinger05}. This source evolution is also used in the case of
GRBs, assuming that GRBs also follow the Star Formation Rate. The energy
density of the observed comic ray spectrum $j_E$ is then used to normalize the
calculated Monte Carlo spectra
\begin{eqnarray}
j_E&:=&\int_{E_{\min}=10^{18.5}~eV}\frac{dN}{dE}\,E\,dE\nonumber\\
&\approx&10^{-7}\diffunits\,.\nonumber
\end{eqnarray}

\begin{figure}[h!]
\begin{center}
\includegraphics[width=7.5cm]{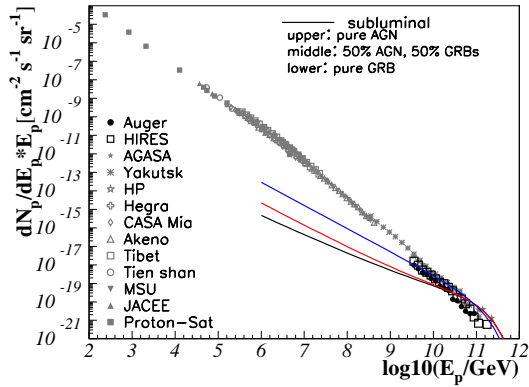}
\caption{The diffuse energy spectrum of relativistic, subluminal sources,
  compared to the total diffuse spectrum.}
\label{diffuse_spectrum}
\end{center}
\end{figure}

The final result is shown
in Fig.~\ref{diffuse_spectrum}. The upper line represents a pure AGN spectrum,
$\alpha=1$, the middle line represents a mixture of $50\%$ AGN and $50\%$ GRBs,
$\alpha=0.5$, and the lower line is a pure GRB spectrum, $\alpha=0$. Since
the cosmic ray spectrum above $E=10^{18.5}$~eV needs to be explained by
extragalactic sources, the best fit results from a pure AGN spectrum. With a
significant contribution, the spectrum becomes too flat.

\section{Conclusions}

In this work we presented Monte Carlo simulation studies on the acceleration of test
particles in relativistic, subluminal shock environments. Source
candidates were applied, such as Active Galactic Nuclei with mild relativistic shocks of 
boost factors of $\Gamma\approx 10$ and Gamma Ray Bursts with highly-relativistic shocks,
$100<\Gamma<1000$.\\
The resulting particle spectra from those shocks have been
investigated in the context of varying shock boost factor $\Gamma$ versus the shock obliquity, 
i.~e.~the inclination angle between the shock normal and the magnetic field, $\psi$. For
three angles, $\psi=23^{\circ},\,33^{\circ},\,43^{\circ}$, shocks were
simulated with a series of different boost factors,
$\Gamma=10,\,20,\,30,\,100,\,300,\,500,\,700,\,900,\,1000$. While the result
remained constant for the different inclination angles, a linear behavior
between the spectral index $\alpha_p$ and the boost factor $\Gamma$ was found,
leading to spectra of $\alpha_p\sim 2.2$ for mildly-relativistic shocks
($\Gamma\sim 10$), but producing much harder spectra for highly-relativistic
shocks. Specifically, for boost factors of $\Gamma\sim 300-1000$, the spectra can become
as flat as $\alpha_p\sim 1.5$.\\
Concerning the observed cosmic ray spectrum at the highest energies, our study shows that one cannot
explain those high energies using solely a contribution from GRBs. On the other hand, a pure AGN spectrum 
fits the spectrum well.

\section{Acknowledgments}
The project is co-funded by the European Social Fund and National
Resources (EPEAEK II) PYTHAGORAS. Part of the work was supported by the DFG
grant LU1495/1-1.

\end{document}